\documentstyle[12pt]{article}
\def\keV{\,{\rm keV}}
\def\MeV{\,{\rm MeV}}
\def\sec{\,{\rm sec}}

\def\rcm{\,{\rm cm}}
\def\km{\,{\rm km}}

\def\eV{{\,\rm eV}}

\def\erg{{\,\rm erg}}
\def\cmm2{{\,\rm cm^{-2}}}
\def\cm2{{\,{\rm cm}^2}}
\def\cmm3{{\,{\rm cm}^{-3}}}
\def\gcmm3{{\,{\rm g\,cm^{-3}}}}

\def\GF{G_{\rm F}}
\def\mnu{m_\nu}
\def\taunu{\tau_\nu}
\def\rdiff{r_{\rm diff}}

\def\Tnu{T_\nu}
\def\nB{n_{\rm B}}
\def\nuN{\sigma_{\nu{\rm N}}}
\def\nunu{\sigma_{\nu\bar\nu}}
\def\neq{n_{\rm eq}}
\def\rns{n_{\rm ns}}
\def\aNN{a_{NN}}

\def\rns{r_{\rm ns}}

\def\Fnu{F_\nu}
\def\rrel{r_{\rm rel}}
\def\Fg{F_\gamma}
\def\Eg{E_\gamma}
\def\Emin{E_{\rm min}}
\def\tmax{t_{\rm max}}

\def\la{\mathrel{\mathpalette\fun <}}
\def\ga{\mathrel{\mathpalette\fun >}}
\def\fun#1#2{\lower3.6pt\vbox{\baselineskip0pt\lineskip.9pt
  \ialign{$\mathsurround=0pt#1\hfil##\hfil$\crcr#2\crcr\sim\crcr}}}
\begin{document}
\pagestyle{empty}
\begin{center}
\bigskip
\rightline{FERMILAB--Pub--94/001-A}
\rightline{astro-ph/9405036}
\rightline{submitted to {\it Physical Review D}}

\vspace{.2in}
{\Large \bf MASSIVE TAU NEUTRINO \\
\bigskip
AND SN 1987A} \\

\vspace{.2in}
G.~Sigl$^{1,2}$ and Michael S. Turner$^{1,2,3}$\\
\vspace{.2in}
{\it $^1$Department of Astronomy \& Astrophysics\\
Enrico Fermi Institute, The University of Chicago, Chicago, IL~~60637-1433}\\

\vspace{0.1in}
{\it $^2$NASA/Fermilab Astrophysics Center\\
Fermi National Accelerator Laboratory, Batavia, IL~~60510-0500}\\

\vspace{0.1in}
{\it $^3$Department of Physics, Enrico Fermi Institute\\
The University of Chicago, Chicago, IL~~60637-1433}\\

\end{center}

\vspace{.3in}

\centerline{\bf ABSTRACT}
\medskip The emission of $\MeV$-mass tau neutrinos from
newly formed neutron stars is considered in a simple, but accurate,
model based upon the diffusion approximation.
The tau-neutrinosphere temperature is found to increase with
mass so that emission of massive tau neutrinos
is not suppressed by the Boltzmann factor previously used,
$(\mnu /T_\nu )^{1.5}\exp(-\mnu/T_{\nu})$, where
$T_{\nu}\sim 4\MeV -8\MeV$.  If the
tau neutrino decays to electron neutrinos, then for short
lifetimes ($\taunu\la10^{-3}\sec$) the location of both the tau and electron
neutrinospheres can be affected, and, for very short lifetimes ($\tau_\nu \la
10^{-6}\sec$) its temperature falls below $1\MeV$, in conflict
with neutrino observations of Supernova 1987A (SN 87A).
Using our results, we revise limits to the mass/lifetime of an $\MeV$-mass
tau neutrino based upon SN 87A.  Our constraints, together with bounds based
upon primordial nucleosynthesis and the laboratory mass limit of around
$30\MeV$, exclude the possibility of a tau neutrino more
massive than $0.4\MeV$ if the dominant decay mode is radiative.
Finally, we speculate on the possible role a $15\MeV -25\MeV$
tau neutrino might play in the supernova explosion itself.

\newpage
\pagestyle{plain}
\setcounter{page}{1}
\newpage

\section{Introduction}

In most extensions of the standard model neutrinos are massive
and unstable.  A massive neutrino can
decay through standard electroweak interactions
into ``visible'' channels such as $\nu \to \nu^\prime \gamma$ or $\nu\to
\nu^\prime \gamma\gamma$, and if it is more massive than about $1\MeV$,
$\nu \to\nu_e e^+e^-$ and $\nu \to\nu_ee^+e^-\gamma$.
It can also decay through new interactions into ``invisible'' channels, e.g.,
$\nu \to \nu^\prime \phi$ or $\nu \to \nu^\prime +
\nu^\prime {\bar\nu}^\prime$, where $\phi$ is a massless
(or very light) boson (e.g., majoron or familon).

Many of the most stringent limits to the radiative decay of a massive neutrino
come from the nondetection of a gamma-ray burst coincident
with the neutrino burst~\cite{nuburst} observed
from Supernova 1987A (SN 87A)~\cite{feilitzsch,chupp,ktsn,bludman,sw,oberauer}
and from the fact that the energy
observed in supernovae in the form of bulk kinetic energy
of the expanding envelope and light (about $10^{51}\erg$ in total)
are far less than that carried away by a neutrino species
(about one third of the the binding energy of a neutron star or
$10^{53}\erg$) \cite{snl}.  These limits are based on neutrino
fluxes calculated within
the framework of standard supernova theory where it is assumed
that all three standard model neutrino species are
massless and stable.  The upper
limits to the masses of the electron and muon neutrinos, about
$7\eV$ and $250\keV$ respectively, are
much less than the temperature of their neutrinospheres,
on the order of $4\MeV$ and $8\MeV$ respectively, and so
this is certainly a good
approximation.  The current laboratory limit to the tau-neutrino
mass is slightly more than $30\MeV$ \cite{labmass} and so
such an approximation is not necessarily
valid for the tau neutrino.  Previously, the
mass dependence of the tau-neutrino energy flux has been obtained
by simply multiplying the energy flux of a massless neutrino species by a
Boltzmann factor, implicitly assuming that the temperature of the
tau neutrinosphere does not depend its mass.

Because of the importance of the SN 87A constraints to the
tau-neutrino \cite{importance}, we felt it worth considering
more carefully the effect of finite mass and lifetime on the energy
fluence of tau neutrinos.  To this end, in Sec. 2 we
develop a simple model of the transport of $\MeV$-mass tau neutrinos
based upon the diffusion approximation.  In Sec. 3 we use
the tau-neutrino energy fluence calculated in this model to
re-derive the constraints that follow from observations of SN 87A.

\section{A Simple Model for Tau-neutrino Transport}
\subsection{Basic considerations}

The energy transport within, and the cooling of, a newly born neutron star
is predominantly due to neutrinos \cite{burrows}.  Because the
neutrino mean free path is very short in a neutron star,
it takes neutrinos a few seconds to diffuse from the center of the
star to the surface.  In considering
the transport of tau and mu neutrinos the most important
processes are elastic, neutral-current scattering off
nucleons, $\nu N \to \nu N$, and pair creation/annihilation
processes involving nucleons and $e^\pm$ pairs, $e^\pm \leftrightarrow
\nu\bar\nu$ and $NN\leftrightarrow NN+\nu\bar\nu$.  For electron
neutrinos charged-current processes involving nucleons and
electrons/positrons are also important.

We define the neutrinosphere to be the surface
beyond which neutrino number is effectively conserved, i.e.,
pair creation/annihilation processes
have become ineffective and chemical equilibrium ceases to
be maintained.  Because the elastic-scattering cross section
is significantly larger than the corresponding
pair creation/annihilation cross sections mu and tau neutrinos
continue to scatter off nucleons well beyond the neutrinosphere.
However, elastic scatterings conserve neutrino number and
approximately conserve neutrino energy, and so
they have little effect on the neutrino energy flux.  Thus,
the flux of mu and tau neutrinos is determined by the conditions at
the neutrinosphere.

In passing, we note that the situation for electron neutrinos
and antineutrinos is different since they can scatter inelastically
through charged-current interactions, e.g., $\nu_e + e^\pm
\to \nu_e + e^\pm$, $\nu_e + n \to e^- + p$, or ${\bar\nu}_e +
p \to e^+ + n$.  The electron neutrinosphere is very close to the mu/tau
neutrinosphere, and in the massless limit the numbers of
electron, mu, and tau neutrinos radiated are essentially
equal.  However, charged-current, inelastic scatterings that take
place beyond the neutrinosphere serve to lower
the kinetic temperature of electron neutrinos, bringing
it closer to the local temperature, ultimately to around $4\MeV$.
This explains an apparent paradox:  namely,
how electron neutrinos can be characterized by a temperature
that is about a factor of two smaller than mu/tau neutrinos
and yet have a fluence that is about the same, rather than a
factor of $(T_{\nu_e}/T_\nu )^3 \sim 1/8$ smaller.

We shall use a simple model based the
diffusion approximation to calculate the
tau-neutrino energy flux; as mentioned above, diffusion is
a very good approximation at the tau neutrinosphere.  The location of the
tau neutrinosphere involves an interplay between scatterings,
pair creation/annihilations, and the local physical conditions
(temperature and density).  We assume that the underlying physical
conditions in the nascent neutron star are the same as those in the
standard case (three massless neutrino species).  Since, energy transport by
tau neutrinos only accounts for about one third of the total
energy transport (and doesn't change radically for the masses
of interest) this assumption is well motivated.
Finally, in actually solving for the flux of tau neutrinos
from the neutrinosphere we use a technique
akin to the freeze-out approximation used in
relic-abundance calculations in the early Universe;
by comparison to direct numerical integration of the transport
equations we find it to be a very good, time saving approximation.

In the massless limit ($m_\nu \ll T_\nu$) we find that the
temperature of the tau neutrinosphere is $7.5\MeV$, in agreement
with standard calculations \cite{burrows}.
Further, our results are very insensitive to the input
parameters (cross sections, model for physical conditions,
and so on), indicating that our various approximations have little
effect on the energy flux of tau neutrinos we calculate.
Both these facts give us confidence that our model provides reliable results.

\subsection{Interactions and physical conditions}

In the diffusion approximation the flux of massive
tau neutrinos $\phi=-D\nabla n$ where $n$ is
the neutrino number density and $D=\lambda v/3$ is the diffusion
coefficient, given by the product of the
neutrino mean free path $\lambda$ and velocity $v$ divided
by three. Throughout,
all energy-dependent quantities are understood to be
averaged over a thermal distribution.  Inside the neutrinosphere
the temperature is given by the local temperature and outside
the neutrinosphere it is taken to be constant and equal
to the temperature of the neutrinosphere.

In general, we approximate thermally averaged quantities
by a functional form that extrapolates correctly to the
ultrarelativistic and nonrelativistic limits; for example,
\begin{equation} \label{vapp}
  v \approx \left({3T\over\mnu+3T}\right)^{1/2}\,.
\end{equation}
Our results are sufficiently robust that the fact that quantities
do not have the correct behaviour in the semirelativistic
regime is of no consequence.

The diffusion coefficient is given by
\begin{equation}\label{Diff}
  D(r) = {v\over3\nB\nuN}\,,
\end{equation}
where $\nB$ is the local baryon number density and $\nuN$ is the cross
section for elastic scattering of massive tau neutrinos
on nucleons (in a hot, nascent neutron star baryons exist
in the form of free nucleons rather than nuclei; see Ref.~\cite{burrows}).
We approximate the elastic-scattering cross section by \cite{bruenn}
\begin{eqnarray}
  v\nuN \approx {\GF^2\over\pi}\left[9T^2+\mnu^2\right]\quad
  \hbox{Dirac}\,,\nonumber\\
  \label{nuNeq}
  v\nuN\approx {\GF^2\over\pi}\left[9T^2+5\mnu T/2\right]\quad
  \hbox{Majorana}\,,
\end{eqnarray}
depending on whether the massive tau neutrino is of the Dirac or of
the Majorana type.

We parameterize the physical conditions in the nascent neutron star during
its early, very hot phase (first $5\sec -10\sec$) by
\begin{equation}\label{mat}
  \nB(r)=n_0\left({r_0\over r}\right)^k\,,\quad
  T (r)=T_0\left({\nB(r)\over n_0}\right)^{1/3}\,,
\end{equation}
where spherical symmetry has been assumed and
for the parameters $r_0$, $n_0$, $T_0$ and $k$ we choose
\begin{equation}
  r_0=30\km\,,\quad n_0=5.97\times10^{35}\cmm3\,,\quad
  T_0=5\MeV\,,\quad k=5\,;
\end{equation}
see e.g., see Ref.~\cite{burrows}.  Our results are insensitive
to the precise parameters chosen.

Deep inside a hot neutron star pair creation/annihilation processes
involving nucleon neutrino-pair brems\-strah\-lung
dominates those involving $e^\pm$ pairs because
of its strong density dependence and the fact that electrons
are very degenerate.  However, in the important region for
our calculation, near the neutrinosphere, the two processes
are of comparable importance.  We write the total cross section
for massive neutrino pair annihilation as their sum,
\begin{eqnarray}
  \nunu v \approx {\GF^2\aNN\nB^2T^{1/2}/m^{5/2}_N\over
  9T^2 + 4\mnu^2/(1+\pi\mnu/2T)^{1/2}}+ 0.83 {\GF^2\over\pi}
  \left[ 9T^2+{5\over8}\mnu^2\right]\quad\hbox{Dirac}
  \,,\nonumber\\
  \nunu v \approx {\GF^2\aNN\nB^2T^{1/2}/m^{5/2}_N\over
  9T^2+ 4\mnu^2 / (1+\pi\mnu/2T)^{1/2}}+0.83 {\GF^2\over\pi}
  \left[ 9T^2+{5\over2}\mnu T\right]\quad\hbox{Majorana}
  \,,\label{nunueq}
\end{eqnarray}
where the numerical constant $\aNN\sim 2.93\times10^3$.
The cross section for the $e^\pm$ process is from
Ref.~\cite{kt} and neglects the fact that electrons are
mildly degenerate; the cross section for the nucleon bremsstrahlung process
is from Refs.~\cite{turner,raffelt} and is computed
in the dilute-medium approximation which is justified for the
highly subnuclear densities around the neutrinosphere.

We also allow for tau-neutrino decays.  The rate of neutrino
decay and inverse decay is controlled by the neutrino lifetime
and energy, with the rate for both processes is reduced
by the Lorentz $\gamma$ factor in the usual way.  We take $22.5\MeV$ to
be the typical energy for tau neutrinos and approximate $\gamma$ by
\begin{equation}\label{gamm}
  \gamma\approx 1+{22.5\MeV\over\mnu} \,.
\end{equation}
In deriving the transport equation we assume that the daughter products
of tau-neutrino decays are in thermal equilibrium
so that detailed balance can be used to supply
the inverse-decay rate.  This assumption is certainly
justified if the decay products are $\nu_e$'s, $\nu_\mu$'s,
$e^\pm$ pairs, or photons and applies to many of the
decay processes of interest.

Finally, for the equilibrium number density of a massive tau neutrino we take
\begin{equation}  \label{neqm}
  \neq \approx \left[{3\zeta (3) \over 4\pi^2}T^3(r)+
  \left({\mnu T(r)\over2\pi}\right)^{3/2}\right]\exp\left(
  -{\mnu\over T(r)}\right)\,,
\end{equation}
which extrapolates to the correct form in the ultrarelativistic
and nonrelativistic limits.

\subsection{Neutrino transport equation}

The neutrino transport equation follows from
adding pair annihilation/creation and decay/inverse decay to the continuity
equation for the neutrino number density
\begin{equation}\label{conti}
  \dot n+\nabla\phi = -\nunu v (n^2-\neq^2)-{(n-\neq )\over\gamma\taunu}
  \,,
\end{equation}
where $\phi$ is the neutrino flux and we
have used detailed balance to relate pair creation to pair
annihilation and inverse decay to decay.

The use of the diffusion approximation and the
assumption of spherical symmetry and stationarity ($\dot n=0$)
allows the transport equation to be written as an ordinary
second-order, nonlinear differential equation
\begin{equation}\label{eqeff}
  Dn^{\prime\prime}+\left(D^\prime+{2\over r}D\right)n^\prime=
  \nunu v (n^2-\neq^2)+{(n-\neq ) \over\gamma\taunu}\,,
\end{equation}
where prime denotes derivative with respect to $r$.

To solve Eq.~(\ref{eqeff}) two boundary conditions must be specified.
To arrive at the first we note that the diffusion
approximation breaks down around the
surface of last scattering ($r=\rdiff$) where neutrinos begin to
stream freely out of the star. For $r\gg \rdiff$ the flux is
therefore given by $\phi\approx nv/2$ where the
factor of two comes from averaging over the angle between the
propagation direction and the surface normal.
On the other hand, where the diffusion approximation
applies ($r\ll \rdiff$) we have $\phi=-D\nabla n$.
Matching these two relations leads to the first boundary condition
\begin{equation}
  -D\nabla n\left\vert_{\rdiff}=(nv /2)\right\vert_{\rdiff}\,.\label{b1}
\end{equation}

Now for the second; well inside the neutrinosphere
the number density of tau neutrinos is very close to its
equilibrium value because annihilation/creation interactions
are occurring rapidly.  We can therefore linearize
Eq.~(\ref{eqeff}) by setting $n=\neq$ in terms that are of
zeroth order in the small quantity $(n-\neq )$ and solve for $n$
to first order in the deviation from $\neq$.  By so doing
and solving for $n (r)$ well inside the neutrinosphere
we obtain our second boundary condition and Eq.~(\ref{eqeff})
can be integrated outward from this inner boundary.
(By varying the position of the inner boundary we have shown that our results
do not depend significantly on it, so
long as it lies considerably inside the neutrinosphere.)

\subsection{The tau neutrinosphere}

While we have solved the neutrino transport equation
for the flux of massive tau neutrinos by direct numerical
integration, we have found
that an approximation based upon the ``freeze out'' of the
number of tau neutrinos provides a good approximation
that requires far less computational time.  In particular,
the tau-neutrino energy fluence calculated agrees within about 50\%.
Given that we are using an approximate model for the nascent
neutron star, this accuracy seems adequate.

Recall, at the neutrinosphere the actual neutrino number
density begins to differ from its equilibrium value because
the mean free path for neutrino pair annihilation/creation
becomes too large to maintain equilibrium.\footnote{The analogy with the
freeze out of a massive particle species in the
early Universe is a good one, and the approximate
technique we use to solve the transport equation is
very similar to the freeze-out approximation used in
relic-particle calculations; see Ref.~\cite{kt}.}
To find the neutrinosphere we algebraically solve Eq.~(\ref{eqeff})
for the radius $\rns$ where $n-\neq$ becomes of order unity
by setting $n-\neq = \neq$ and $n^2-\neq^2 =
\neq^2$:\footnote{As it turns out, there is a
small mass range where this procedure
does not work well because the term on the left-hand side
becomes zero.   To insure that $\rns$ is a smooth function
of $\mnu$ and $\taunu$ we actually take the maximum of
this term and 0.1 times the modulus of an analogous term with a
minus sign between the second and the first order derivatives
which could otherwise cancel each other.}
\begin{equation}\label{eqrns}
  \left\vert D\neq^{\prime\prime}+\left(D^\prime+{2\over r}D
  \right)\neq^\prime\right\vert_{\rns}=
  \left[v\nunu\neq^2+{\neq\over\gamma\tau}\right]_{\rns}
  \,.
\end{equation}
{}From this, we approximate the power radiated in massive tau-neutrinos,
\begin{equation}
  P_\nu (m_\nu \not= 0) \approx [\mnu+3\Tnu]
  4\pi\rns^2[D\neq^\prime]_{\rns}\,,\label{flux}
\end{equation}
where $\Tnu\equiv T(\rns)$ is the temperature at the tau neutrinosphere.
While the power agrees well with the
standard results in the massless limit, we shall actually
use the ratio of the power for a massive tau neutrino to that
of a massless species and the standard results for a massless
neutrino species to obtain absolute results for the
energy fluence (in any case, the total energy radiated
depends upon the cooling time of the hot neutron star).
The number fluence of and energy carried by a massless neutrino species
(both neutrinos and antineutrinos) are
about $1.4\times 10^{10}\rcm^{-2}$ and $10^{53}\erg$ respectively.

Our results for the tau neutrinosphere temperature and tau-neutrino
energy fluence are shown in Figs.~1 and 2 as a function of tau neutrino mass
and lifetime.  In the limit of long lifetime ($t\ga 10^{-2}\sec$)
the tau neutrinosphere temperature remains constant at about
$7.5\MeV$ until a mass of about $10\MeV$, after which it rises
steadily, reaching around $14\MeV$ for a mass of $100\MeV$.
The reason for the rise is simple:  As the mass increases
beyond $10\MeV$ the tau-neutrino equilibrium abundance at the
``massless neutrinosphere'' decreases exponentially; pair-annihilation
processes vary as the neutrino abundance squared
and cannot reduce the neutrino number density to its small
equilibrium value for large tau-neutrino masses; thus, the
tau neutrinosphere moves inward,
to a higher temperature where the abundance is larger.
The net result is that the tau-neutrino energy fluence relative
to a massless neutrino species decreases far more slowly than
it would if one (incorrectly) assumes that the neutrinosphere
temperature is independent of neutrino mass; see Fig.~2.

In Figs.~1 and 2 the effect of the tau-neutrino lifetime
upon the neutrinosphere is also shown.  For lifetimes greater
than about $10^{-2}\sec$ there is little effect; as the lifetime
is decreased, which corresponds to increasing the rate of
decays and inverse decays, the tau-neutrinosphere temperature
decreases.  The reason is simple:  Inverse decays and decays
become very effective at maintaining chemical
equilibrium, further and further out.  In fact, they are so
effective that for tau-neutrino lifetimes shorter than about
$10^{-3}\sec$ the tau neutrinosphere temperature drops
below $7\MeV$, and presumably, the electron
neutrinosphere temperature is lowered too.  For very
short lifetimes the effects of decays and inverse decays
are so severe that the background model is likely to be affected too,
calling into question its validity.

Finally, a brief, but important, comment on the robustness
of our results.  Consider the tau-neutrinosphere as defined
above.  Its location depends upon the model of the ambient
conditions as well as the rates for elastic scattering and
pair-creation/annihilation.  Varying the model for the
ambient conditions, e.g., $k$ and $T_0$, affects the tau-neutrinosphere
temperature by only a few tenths of an $\MeV$.  Likewise,
it depends very little upon the overall normalization of
the scattering and annihilation rates adopted:  In particular,
it is the product of the rates that determines $T_\nu$, and
changing that product by a factor of ten only changes
the tau-neutrinosphere temperature by about $1\MeV$ and the
energy flux by less than a factor of four for long-lived (i.e.
$\taunu\ga0.1\sec$) tau neutrinos.  Insensitivity
to input parameters is also apparent in
the small differences in the results for Dirac and Majorana
neutrinos, where the product of the rates differs by a
factor $0.1(\mnu/\Tnu)^2$ for $\mnu\ga20\MeV$.

\subsection{Wrong-helicity neutrinos}

A massive Dirac neutrino has two
additional helicity states, $\nu_R$ and ${\bar\nu}_L$,
which in the absence of new interactions only interact
by virtue of the mismatch between chirality and helicity.
For every ordinary process, e.g., $\nu_L +N \to \nu_L+N$,
there is a spin-flip analog, here, $\nu_L + N \to \nu_R +N$,
which is suppressed by a factor of $(m_\nu /2E_\nu)^2$.

For masses much less than $1\MeV$ the suppression factor
is very large and the mean-free path
of a wrong-helicity neutrino is much larger than the radius
of a neutron star.  Wrong-helicity neutrinos that are produced
simply stream out, tending to accelerate the cooling of
a hot neutron star, e.g., the one associated with SN 87A, and thereby
shortening the duration of the predicted neutrino signal.
The duration of the neutrino burst observed by KII and
IMB has been used to exclude a Dirac neutrino of mass between about
$10\keV$ and $1\MeV$ \cite{dirac}.

For masses much greater than $1\MeV$ the suppression factor is
not significant, and wrong-helicity neutrinos become trapped and are radiated
from a ``wrong-helicity'' neutrinosphere.  Using our diffusion model
we have estimated the location of the wrong-helicity neutrinosphere.
(The dominant interactions for both scattering
and creation/destruction are spin-flip neutrino-nucleon scatterings;
see Ref.~\cite{turner}.)  For a mass of $1\MeV$ the wrong-helicity
neutrinosphere temperature is about $40\MeV$, and wrong-helicity
neutrinos will quickly rob the core of the hot neutron star
of most of its thermal reserves.  As before, this is excluded by the
KII and IMB data.  As the mass increases the temperature of
the wrong-helicity neutrinosphere decreases, reaching a
temperature comparable to that of the proper-helicity neutrinosphere
for $m_\nu = 5\MeV$.  For $m_\nu \gg 5\MeV$ both neutrinospheres
coincide, thereby roughly doubling the energy flux
(compared to a Majorana neutrino).

The two neutrinospheres are coupled (through spin-flip
interactions); to properly take account of both helicity states of an MeV
Dirac neutrino is beyond the scope of this study.  Since we are
only interested in neutrino masses greater than a few MeV,
where wrong-helicity neutrinos carry off only about as
much energy as proper-helicity neutrinos, to be conservative
we simply ignore the energy carried off by wrong-helicity
neutrinos.  (Doubling our energy fluxes would change our results very little.)

\section{Constraints Based Upon SN 87A}

In this Section we re-examine the important mass-lifetime limits to an
MeV-mass tau neutrino based on SN 87A using the more accurate tau-neutrino
fluence computed here.  In general,
our limits are significantly more stringent as the tau-neutrino
energy fluence is larger than had previously been assumed.
The limits we discuss are based upon three observations:
(i) the total ``visible''
energy, the kinetic energy of the expanding shell of
matter (about $10^{51}\erg$) and the
optical light output (about $10^{49}\erg$), was less than
$10^{51}\erg$ \cite{arnett}; (ii) the gamma-ray fluence
around the time when the neutrinos were detected
was very small compared to the enormous neutrino fluence, $\Fnu\sim
10^{10}\rcm^{-2}$ \cite{chupp}; and (iii) 19 neutrino events were
detected by the KII and IMB detectors \cite{nuburst},
consistent with antielectron neutrinos emitted with
a kinetic temperature of about
$4\MeV$ \cite{burrows}.  In deriving our
mass-lifetime limits for a tau neutrino of mass
greater than a few $\MeV$, we follow closely the treatments
in Refs.~\cite{oberauer,uc} and also assume that the decay
rate is dominated by visible channels (i.e., daughter products
include photons or $e^\pm$ pairs).

\subsection{Decays inside the progenitor}

The progenitor of SN 87A had a size of about $3\times 10^{12}\rcm$
\cite{arnett}.  For tau-neutrino lifetimes of the order $100\sec$ or smaller
an appreciable fraction of the tau neutrinos emitted
decay inside the progenitor and deposit energy.  Since both
$e^\pm$ pairs and photons have small mean free paths inside
the progenitor and thus should deposit all their
energy within the envelope of the progenitor,
our arguments do not depend whether the daughter
products are $e^\pm$ pairs or photons.
The fraction of tau neutrinos that decay inside the progenitor
\begin{equation}
f_{\rm inside} = 1-\exp(-t_{\rm inside}/\tau_L ),
\end{equation}
where $t_{\rm inside} \approx 100\sec /v$ is the time it
takes a tau neutrino to escape the progenitor and
$\tau_L = \gamma \tau_\nu$ is the lifetime in the
progenitor rest frame.

We demand that the energy deposited inside the progenitor $E_{\rm inside}$
be less than about $10^{51}\erg$ to be consistent with
the observations of SN 87A (and other type II supernovae):
\begin{equation}\label{SNL}
        E_{\rm inside} \approx f_{\rm inside} 10^{53}\erg\,
        \left( {P_\nu (m_\nu \not= 0)\over P_\nu (m_\nu = 0)}
        \right) \la 10^{51}\erg .
\end{equation}
The region of the mass-lifetime plane excluded by this
consideration is shown in Fig.~3 (labeled SNL).

\subsection{Decays outside the progenitor}

The differential gamma-ray fluence at Earth due to
tau-neutrino decays via the channel $\nu_\tau\to\nu_e\gamma$
was calculated in Ref.~\cite{oberauer}.  As modified to our
approximations it is
\begin{equation}\label{gam}
  {d\Fg\over d\Eg}={B_\gamma\Fnu\Eg\over\Tnu\mnu\taunu}
  \left({\rns\over\rrel}\right)^2\int_0^{\tmax} dt
  \exp(-2\Eg t/\mnu\taunu)
  \left(1+{\Emin\over\Tnu}\right)\exp(-\Emin/\Tnu )\,,
\end{equation}
where $\rns$ and $\Tnu$ are the radius and temperature of
the tau neutrinosphere respectively, $\rrel$ is the radius
of the light neutrinosphere, $\Emin$ is the maximum of $\Eg$ and
$$\mnu\left[1+(100\sec\cdot\mnu/2\Eg t)^{1/2}\right],$$
and $\tmax=223.2\sec$ is the time interval over which
the authors of Ref.~\cite{oberauer} analysed the data recorded
by the Gamma-Ray Spectrometer on board the Solar Maximum
Mission satellite. Eq.~(\ref{gam}) differs from the original expression
in Ref.~\cite{oberauer} in two respects: First, the neutrinosphere
temperature $\Tnu$ depends upon tau-neutrino mass; and second,
the tau-neutrino fluence is must be modified by the
geometrical factor $(\rns/\rrel)^2$ relative to the light
neutrino fluence $\Fnu=1.4\times10^{10}\rcm^{-2}$.

In this case we are only interested in decays that produce
a gamma ray.  However, for $\MeV$-mass tau neutrinos the
mode $\nu_\tau \rightarrow
\nu +e^\pm$ is kinematically allowed and likely dominates
the photonic mode.  Assuming it proceeds via standard
electroweak interactions,
\begin{equation}
\tau_\nu = {192\pi^3 \over G_F^2 m_\nu^5 \sin^2\theta
\cos^2\theta} \simeq {1 \sec \over \sin^2 2\theta (m_\nu /10\MeV )^5}\,,
\end{equation}
where $\theta$ is the $\nu_\tau - \nu_e$ mixing angle.
To wit, we have taken the factor
$B_\gamma \sim 10^{-3}$, corresponding to the
fraction of decays expected to have a bremsstrahlung photon
(also see Ref.~\cite{mohapatra}).  Integrating Eq.~(\ref{gam})
over the suitable energy bands and comparing with the corresponding $3\sigma$
fluence limits reported in Ref.~\cite{oberauer} leads to the
excluded region denoted SMM in Fig.~3 \cite{pvo}.

\subsection{SN 87A neutrinos}

The nineteen neutrino events recorded by the KII and IMB detectors
confirmed the standard model of the cooling of
a nascent neutron star and had an energy distribution
consistent with a temperature of about $4\MeV$ \cite{burrows}.
As noted earlier and illustrated in Figs.~2b and 2c, for
very short lifetimes ($\tau_\nu \la 10^{-6}\sec$),
decays and inverse decays become so
potent that they move the tau and electron neutrinospheres
outward to a region where the temperature is around $1\MeV$ or less.
Clearly this is inconsistent with the KII and IMB data.
However, we hasten to add that our model for the neutrinosphere is
not self consistent under such extreme conditions.
More importantly, lifetimes for the modes $\nu_\tau \to
\nu_e+e^\pm$ or $\nu_\tau \to \nu +\gamma$
shorter than about $1\sec$ are excluded by
data from the Big European Bubble Chamber (BEBC) experiment
\cite{short} (shown in Fig.~3).  Unless new interactions beyond the
standard model can give rise to a very short lifetime
for a nonelectromagnetic mode where all the daughter products
would be in thermal equilibrium in a hot neutron star,
it would seem that our results for very short lifetimes
are without application.

In passing we mention that if tau neutrinos decay beyond the
electron neutrinosphere, corresponding to $\tau_\nu \ga 10^{-3}\sec$,
there is an additional population of antielectron neutrinos,
which for $m_\nu \la 10\MeV$ is of comparable importance.
It appears unlikely that our understanding of SN 87A and the KII/IMB data are
good enough to exclude this possibility.  However, because
$\MeV$-mass neutrinos are only semi-relativistic, these additional
events would be characterized by a time delay/spread on
the order of the tau-neutrino lifetime itself and thus
could perhaps be excluded (or confirmed) on that basis.
(In fact, it could be wildly speculated that a tau neutrino
with a lifetime of seconds and a mass of $1\MeV - 10\MeV$
could be responsible for the second cluster of KII events.)

\subsection{Residual annihilations}

Finally, we would like to draw attention to an interesting
aspect of the behaviour of the
numerical solutions of Eq.~(\ref{eqeff}) in the long-lifetime limit.
In Figs.~4 we show the
solution and the equilibrium distribution for the
massless case and for $\mnu=30\MeV$.  The departure from equilibrium
near the neutrinosphere radius is clearly visible. The kinks
mark the surface of last scattering ($r=\rdiff$) where the
boundary condition Eq.~(\ref{b1}) applies.
In the massless case the actual tau-neutrino number density
``undershoots'' the equilibrium density in the diffusion
region, whereas in the massive case it ``overshoots.''
The ``overshoot effect'' suggests that a massive
neutrino could be effective in depositing energy into the
outer core region by residual annihilations and thus
could help power the supernova explosion itself.  (At
present, it is not understood precisely how the explosion
is powered \cite{burrows,arnett}).

To study this further we calculated the energy
deposited in the region between the light neutrinosphere and
the outer boundary of the diffusion region.  (Inside
the light neutrinosphere any energy deposited will be
radiated in electron/muon neutrinos, and beyond the
diffusion region annihilations are totally ineffective
because of the free streaming of neutrinos).
We approximate this energy by
\begin{equation}\label{Pnu}
  E_{\nu\bar\nu} \sim (\mnu+3\Tnu)\int_{\rrel}^{\rdiff}4\pi r^2 dr
  \nunu v\left(n^2-\neq^2\right)\,,
\end{equation}
where $\nunu v$ is evaluated at the tau neutrinosphere and $n(r)$
is the analytical solution of Eq.~(\ref{eqeff}) between $\rns$
and $\rdiff$ for vanishing right-hand side and
$D(r)=D(\rns)(r/\rns)^k$.  In Fig.~5 we show the ratio of the
energy deposited to the energy carried off by a massless
neutrino species (about $10^{53}\erg$).  For tau-neutrino masses
between about $15\MeV$ and $25\MeV$ the energy deposited
is about $10^{51}\erg$, which is comparable to the
energy seen in the kinetic motion of the explosion and thus
could have an important effect on the supernova explosion---perhaps
solving the riddle of the explosion itself.

\section{Concluding Remarks}

We have studied the transport of energy by $\MeV$-mass tau neutrinos
during the early cooling phase of a hot neutron star.
By means of a simple, but accurate, diffusion model we have
computed the energy fluence and neutrinosphere temperature.
Because the neutrinosphere temperature increases with tau-neutrino
mass the energy carried off by massive tau neutrinos does
not decrease nearly as rapidly as it would if
the tau-neutrinosphere temperature were independent
of mass, as previously assumed.  (We mention that our method
could easily be generalized to calculate the flux of hypothetical,
massive particles with interactions whose strength is roughly
weak---e.g., new particles predicted in supersymmetric models---if
the motivation should arise.)

Using these new results we have re-derived the mass-lifetime
bounds that follow from SN 87A.  Based upon SMM gamma-ray data
we exclude a massive tau neutrino of lifetime shorter than
about $10^8\sec$ whose decays are radiative.  This, taken with the
recent bounds based upon primordial nucleosynthesis \cite{uc,osu},
excludes a tau neutrino more massive than about $0.4\MeV$ regardless
of its lifetime, provided only that its decays are radiative.
In addition, a Dirac tau neutrino of mass from about $10\keV$
to $1\MeV$ is excluded on the basis of the fact that wrong-helicity
tau-neutrinos produced deep in the core would have carried off
too much energy, leading to an unacceptably short burst
of antielectron neutrinos \cite{dirac}.
Finally, we speculate that a tau neutrino of mass $15\MeV -
25\MeV$ could play an important role in the supernova
explosion process itself by virtue of the energy its
annihilations deposit just outside the tau neutrinosphere.

\section*{Acknowledgments}

We thank Adam Burrows, Scott Dodelson and David Seckel for helpful comments.
This work was supported in part by the Department of Energy
(at Chicago and at Fermilab), by the NASA through grant
NAGW-2381 (at Fermilab) and by the Alexander-von-Humboldt Foundation.

\newpage

\section*{Figure Captions}
\bigskip

\noindent{\bf Figure 1:}  Tau-neutrinosphere temperature as
a function of tau-neutrino mass:  (a)  Long-lifetime limit
(upper curves) and $\taunu = 10^{-5}\sec$ (lower curves)
for Dirac (solid) and Majorana (broken) mass;
(b) Contours in the mass-lifetime plane for Dirac mass; and
(c) Contours in the mass-lifetime plan for Majorana mass.
Contours are in steps of $0.5\MeV$ beginning with $7.5\MeV$ in
the upper left corner. The temperature increases to the right
and decreases to the bottom.

\medskip
\noindent{\bf Figure 2:}  Massive tau-neutrino energy flux relative
to a massless tau neutrino:  (a)  Long-lifetime limit (upper curves)
and $\taunu = 10^{-5}\sec$ (lower curves) for Dirac (solid)
and Majorana (broken) mass compared to the naive Boltzmann
suppression factor for mass-independent neutrinosphere
temperature $\Tnu=7.5\MeV $ (dotted); (b) Contours in the mass-lifetime
plane for Dirac mass; and (c) Contours in the mass-lifetime plane
for Majorana mass. Flux contours decrease by a factor of
10 (per contour) from top left to the lower right.

\medskip
\noindent{\bf Figure 3:}  Excluded regions of the tau-neutrino
mass-lifetime plane:  (a) Dirac mass; and (b) Majorana mass.
Forbidden regions are on the same side as the label; SNL is the
limit based upon decays inside the progenitor; SMM is
the limit based upon decays outside the progenitor and
the excluded region is between the curve just above
SMM and the unmarked lower curve; CLEO/ARGUS
refers to the laboratory mass limit; and BEBC refers to the
laboratory lifetime limit discussed in Sec. 3.3.

\medskip
\noindent{\bf Figure 4:}  Numerical solution for tau-neutrino
number density (solid curve) and equilibrium number density (broken curve),
both relative to the neutrino number density at the inner boundary
(see Sec. 2.3): (a) massless tau neutrino;
and (b) $30\MeV$ Dirac tau neutrino.

\medskip
\noindent{\bf Figure 5:}  Energy deposited by residual annihilations
of a tau neutrino as a function of tau-neutrino mass in units of
$10^{53}\erg$; solid curve is for Dirac and broken curve is for Majorana.

\end{document}